\documentclass[aps,pra,notitlepage,superscriptaddress,showpacs,amsmath,amssymb]{revtex4-1}
\usepackage{graphicx}
\pdfoutput=1

\DeclareMathOperator{\re}{Re}

\begin{document}

\title{Creation of two-photon states via interactions between Rydberg atoms
during light storage}

\author{J. Ruseckas}
\affiliation{Institute of Theoretical Physics and Astronomy, Vilnius University,
Saul\.etekio ave.\ 3, LT-10222 Vilnius, Lithuania}

\author{I. A. Yu}
\affiliation{Department of Physics and Frontier Research Center on Fundamental
and Applied Sciences of Matters, National Tsing Hua University, Hsinchu
30013, Taiwan}

\author{G. Juzeli\=unas}
\affiliation{Institute of Theoretical Physics and Astronomy, Vilnius University,
Saul\.etekio ave.\ 3, LT-10222 Vilnius, Lithuania}

\begin{abstract}
We propose a new method to create two-photon states in a controllable
way using interaction between the Rydberg atoms during the storage
and retrieval of slow light. A distinctive feature of the suggested
procedure is that the slow light is stored into a superposition of
two atomic coherences under conditions of electromagnetically induced
transparency (EIT). Interaction between the atoms during the storage
period creates entangled pairs of atoms in a superposition state that
is orthogonal to the initially stored state. Restoring the slow light
from this new atomic state one can produce a two photon state with
a second-order correlation function determined by the atom-atom interaction
and the storage time. Therefore the measurement of the restored light
allows one to probe the atom-atom coupling by optical means with a
sensitivity that can be increased by extending the storage time. As
a realization of this idea we consider a many-body Ramsey-type technique
which involves $\pi/2$ pulses creating a superposition of Rydberg
states at the beginning and the end of the storage period. In that
case the regenerated light is due to the resonance dipole-dipole interaction
between the atoms in the Rydberg states.
\end{abstract}


\maketitle

\section{Introduction}

Generation of photon pairs has a fundamental and technological significance
\cite{Yamamoto1999}, and can be used in quantum measurement and quantum
information transfer \cite{Bouwmeester2000}. Production of non-classical
photon pairs via the electromagnetically induced transparency (EIT)
has been first demonstrated a decade ago \cite{vanderWal2003,Kuzmich2003}.
Subsequently photon pairs with a controllable profile have been created
\cite{Balic2005,Du2008} by employing the slow light in a double-$\Lambda$
system. In related recent developments, nonlinear quantum optics
has been investigated for slow light using Rydberg atoms
\cite{Petrosyan2011,Gorshkov2011,Dudin2012,Dudin2012a,Peyronel2012,Firstenberg2013,Maxwell2013,Li2013,Hofmann2013,Chang2014,He2014}.
Since the van der Waals interaction between the atoms increases with
the principal quantum number as $n^{11}$ , the interaction between
the Rydberg atoms is enhanced by many orders of magnitude compared
to the interaction between atoms in the ground state \cite{Mohapatra2007,Bohlouli2007,Pritchard2010,Beguin2013}.
The interaction brings neighboring Rydberg atoms out of the resonance
destroying the EIT. Consequently the closeby Rydberg atoms absorb
the slow light, so photons become antibunched during propagation
of light through the atomic medium 
\cite{Petrosyan2011,Gorshkov2011,Dudin2012,Dudin2012a,Peyronel2012,Firstenberg2013,Li2013,Chang2014}.
Nonclassical photon or atomic states using Rydberg interactions have also
been investigated in Refs.~\cite{Pritchard2012,Stanojevic2012,Bariani2012}.

Here we propose another way of generation of correlated two-photon
states via storage and retrieval of the slow light in the atomic medium.
Unlike conventional light storage
\cite{Fleischhauer-PRL-2000,Hau-2001,Phillips-PRL-2001,Fleischhauer2002,Juzeliunas2002,Lukin-03,Fleischhauer-05,Hau07Nature,Bloch09PRL},
the probe pulse is now stored in a superposition of two atomic states
involving the Rydberg levels. During a subsequent evolution the atom-atom
interaction produces entangled pairs of atoms in a superposition state
that is orthogonal to the initially stored state. Restoring the slow
light from this new atomic state one can produce two photon states
with the second-order correlation function determined by the atom-atom
interaction and the storage time. Furthermore, measurement of the
second-order correlation function of the restored light allows one
to probe the interaction potential by optical means, with a sensitivity
that can be enhanced by increasing the storage time. Note that the
creation of an atomic superposition and a subsequent retrieval of
light form the orthogonal superposition represents a Ramsey interferometry
\cite{Ramsey1950,Knap2013,Yan2013,Zhang2014,Mukherjee15}. 

Previously regeneration of light from an initially unpopulated coherence
was implemented using an external detuning in a tripod atom-light
coupling scheme \cite{Lee2014} (see also a related work \cite{Jian-Wei-Pan2015PRA}).
In that case the regenerated light remains classical by storing a
classical light. On the other hand, in the current proposal the detuning
is caused by the interaction between closeby Rydberg atoms leading
to regeneration of correlated photon pairs.

As a specific realization of the proposed idea we consider in
this paper a light-matter interaction in an ensemble of atoms characterized
by a ladder scheme of energy levels shown in Fig.~\ref{fig:scheme}(a). After
switching off the control beam, the probe beam is stored in a coherence between
the ground state $g$ and the Rydberg state $s$.  In order to create a
superposition of two Rydberg states and to restore the slow light from an
orthogonal superposition we propose to apply at the beginning and at the end of
the storage the $\pi/2$ optical pulses coupling the Rydberg $s$ and $p$
states.

The paper is organized as follows. In Sec.~\ref{sec:formulation} we present the
proposed setup. In Sec.~\ref{sec:storage} we investigate how the stored atomic
state is changed by the atom-atom interactions.  In Sec.~\ref{sec:retrieval} we
analyze the probe pulse restored from this atomic state.
Section~\ref{sec:concl} summarizes our findings and discusses possible
experimental implementation.

\section{Formulation}
\label{sec:formulation}

\begin{figure}
\includegraphics[width=0.6\textwidth]{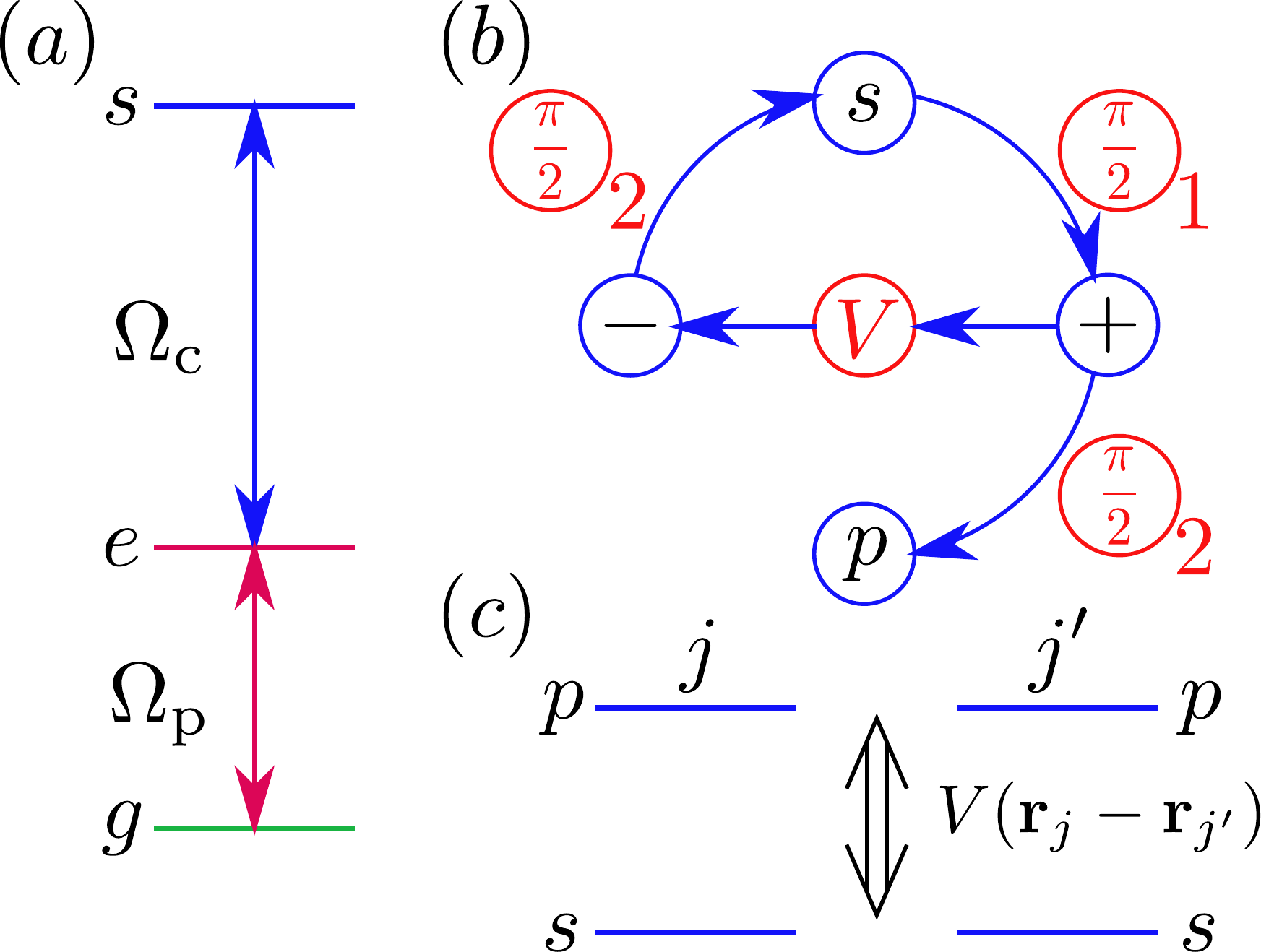}
\caption{(color online). (a) The atomic ground state $|g\rangle$ is coupled to
  the excited state $|e\rangle$ by the probe field $\Omega_{\mathrm{p}}$.  The
  latter $|e\rangle$ is coupled to the Rydberg state $|s\rangle$ by the control
  laser $\Omega_{\mathrm{c}}$. (b) Changes of the atomic states due to the
  action of the first $\pi/2$ pulse, the RDDI potential $V$ and the second
  $\pi/2$ pulse. (c) The RDDI $V=V(|\mathbf{r}_{j}-\mathbf{r}_{j'}|)$ exchanges
  of the $s$ and $p$ Rydberg states between nearby atoms $j$ and $j^{\prime}$.}
\label{fig:scheme}
\end{figure}

We consider a light-matter interaction in an ensemble of atoms characterized by
a ladder scheme of energy levels shown in Fig.~\ref{fig:scheme}(a).  We assume
that the size of the atomic medium is much larger than the optical wavelength.
We include the atomic ground $s$ and excited $p$ states labeled, respectively,
by $|g\rangle$ and $|e\rangle$, as well as Rydberg $s$ and $p$ states denoted
by $|s\rangle$ and $|p\rangle$. The corresponding energies are
$\hbar\omega_{g}$, $\hbar\omega_{e}$, $\hbar\omega_{s}$ and $\hbar\omega_{p}$.
The atoms, initially in the ground state $|g\rangle$, are illuminated by a
probe laser field with a central frequency $\omega_{\mathrm{p}}$.  Additionally
there is a more intense classical control field with a frequency
$\omega_{\mathrm{c}}$.  The probe (control) field resonantly drives the atomic
transition $g\rightarrow e$ ($e\rightarrow s$) with a coupling strength
characterized by a Rabi frequency $\Omega_{\mathrm{p}}$
($\Omega_{\mathrm{c}}$). 

The incident probe field $\Omega_{\mathrm{p}}$ represents a long and flat pulse
of a classical light, such that, except for a short transient period, it can be
considered to be time independent. The second-order correlation function
$g_{\mathrm{in}}^{(2)}(\tau)$ of such an incident light is constant. We are
interested in times $\tau$ corresponding to interatomic distances
$r=v_{g0}\tau$ much larger than the Rydberg blockade radius, so that the
interaction between atoms can be neglected during the propagation of light with
a velocity $v_{g0}\ll c$ in the medium.  In the previous studies on the
nonclassical slow light
\cite{Lukin2001,Petrosyan2011,Gorshkov2011,Peyronel2012,Firstenberg2013} the
strong Rydberg blockade during the propagation of slow light provides photons
antibunched over a length exceeding the blockade radius. In contrast, the
present approach does not rely on the Rydberg blockade, so one can neglect its
effects during the propagation of the slow light.

After switching off the control beam, the probe beam is stored in a coherence
between the ground state $g$ and the Rydberg state $s$.  Subsequently a $\pi/2$
optical pulse is applied that couples the Rydberg $s$ and $p$ states, as shown
in Fig.~\ref{fig:scheme}(b).  This converts the Rydberg $s$ state into a
superposition of the $s$ and $p$ Rydberg states
$|+\rangle=(|s\rangle+|p\rangle)/\sqrt{2}$.  A similar procedure has been
employed in Ref.~\cite{Maxwell2013}, where a single microwave pulse has been
used during the light storage to couple the initial  Rydberg state to a
neighboring internal state.  The medium is then left to evolve freely for a
duration $T$. As we shall see later on, during the storage the resonance
dipole-dipole interaction (RDDI) between the atoms in the $s$ and $p$ Rydberg
states creates \emph{correlated pairs} of atoms $j$ and $j^{\prime}$ in an
initially unpopulated state $|-\rangle=(|s\rangle-|p\rangle)/\sqrt{2}$ with
correlations determined by the RDDI potential
$V(|\mathbf{r}_{j}-\mathbf{r}_{j'}|)$.  Note that high fidelity $\pi/2$
microwave pulses should be of a sufficiently large Rabi frequency which exceeds
the corresponding strength of the dipole-dipole interaction
$\Omega_{\mathrm{rf}}>V(r)$ at relevant interatomic distances $r$.

Just before the retrieval one applies another $\pi/2$ optical pulse coupling
the Rydberg $s$ and $p$ states. This converts the state $|+\rangle$ into the
Rydberg state $|p\rangle$, whereas the state $|-\rangle$ is transferred back
into the Rydberg state $|s\rangle$; see Fig.~\ref{fig:scheme}(b). Such a
procedure represents a Ramsey-type interferometry involving atom-atom
interaction \cite{Knap2013,Yan2013,Zhang2014,Mukherjee15}.  For atoms in the
Rydberg state $|p\rangle$ there is no allowed optical transition to the excited
state $|e\rangle$. Therefore, when restoring the light, atomic excitations in
the $s$ state are converted into probe photons, and the $p$ state excitations
remain in the medium.  Hence no slow light would be regenerated without the
RDDI which converts the internal state $|+\rangle$ into $|-\rangle$ for
neighboring atoms. Restoring the probe beam one produces correlated pairs of
probe photons, like in the parametric downconversion \cite{Loudon2000}.

Applying the rotating wave approximation \cite{Scully1997}, a Hamiltonian for
the atoms coupled with the laser fields reads in the interaction representation 
\begin{equation}
\mathcal{H}=\mathcal{H}_{\mathrm{at-light}}+\mathcal{H}_{\mathrm{SP}}
+\mathcal{H}_{\mathrm{at-at}}\,.\label{eq:ham}
\end{equation}
Here 
\begin{equation}
\mathcal{H}_{\mathrm{at-light}}=-\frac{1}{2}\sum_{j}\left(\Omega_{\mathrm{p}}
\sigma_{eg}^{j}+\Omega_{\mathrm{c}}\sigma_{es}^{j}+\mathrm{H.c.}\right)
\label{eq:ham-at-light}
\end{equation}
is a Hamiltonian for the atom-light coupling,
$\sigma_{ab}^{j}=|a^{j}\rangle\langle b^{j}|$ are (quasi)-spin-flip operators
transferring the $j$th atom from state $b$ to state $a$
\cite{Fleischhauer-05}, with $a$ and $b$ standing for the atomic internal
states $g$, $e$, $s$, and $p$. We have assumed an exact EIT resonance with zero
two- and single-photon detunings:
$\omega_{g}+\omega_{\mathrm{p}}+\omega_{\mathrm{c}}-\omega_{s}=0$ and
$\omega_{g}+\omega_{\mathrm{p}}-\omega_{e}=0$. The term
\begin{equation}
\mathcal{H}_{\mathrm{SP}}=\sum_{j}\Omega_{\mathrm{SP}}(t)(\sigma_{ps}^{j}
+\sigma_{sp}^{j})
\label{eq:ham-S-P}
\end{equation}
describes the coupling between the Rydberg $s$ and $p$ states due to an
external electromagnetic field characterized by a Rabi frequency
$\Omega_{\mathrm{SP}}(t)$. The latter $\Omega_{\mathrm{SP}}(t)$ is composed of
two $\pi/2$ pulses, one applied at the beginning of the storage, another one at
the end of the storage. At the remaining stages (propagation, storage, and
release of light) $\Omega_{\mathrm{SP}}(t)$ is off and hence is to be omitted.
Finally 
\begin{equation}
\mathcal{H}_{\mathrm{at-at}}=\sum_{j\neq j'}V(|\mathbf{r}_{j}
-\mathbf{r}_{j'}|)\sigma_{ps}^{j}\sigma_{sp}^{j'}\label{eq:ham-at-at}
\end{equation}
is the RDDI Hamiltonian leading to exchange of Rydberg states between pairs of
atoms, one of which being in the $s$ state and another in the $p$ state. The
action of RDDI during the light storage is schematically shown in
Fig.~\ref{fig:scheme}(c). The strong RDDI between pairs of Rydberg atoms is
described by a position-dependent strength
$V(|\mathbf{r}_{j}-\mathbf{r}_{j'}|)$, chosen to be real, where
$\mathbf{r}_{j}$ is a position vector of the $j$th atom. The condition $j\neq
j'$ excludes self-interactions in Eq.~(\ref{eq:ham-at-at}). Note that a double
summation over $j$ and $j'$ ensures that both forward and backward resonance
transfer between the states $s$ and $p$ are included in the interaction
Hamiltonian (\ref{eq:ham-at-at}) which is therefore Hermitian. 

\section{Time evolution of atomic state during the storage of probe pulse}
\label{sec:storage}

\subsection{Stored atomic state}

When the control field $\Omega_{\mathrm{c}}$ is on, the slow light made of
dark-state polaritons propagates in the medium with a group velocity
$v_{g0}=\Omega_{\mathrm{c}}^{2}L/\Gamma\alpha$
\cite{Harris-PT-1997,Hau1999,Fleischhauer-PRL-2000,Fleischhauer2002}.  Here
$\alpha$ and $L$ are, respectively, an optical density and a length of the
medium, whereas $\Gamma$ is a decay rate of the excited state $e$. Since the
initial probe field is classical, one can replace atomic spin-flip operators
$\sigma_{ab}^{j}$ by the corresponding density matrix elements
$\rho_{ba}^{j}=\langle\sigma_{ab}^{j}\rangle$.  A duration $T$ of the
subsequent light storage is assumed to be much larger than the propagation time
of slow light in the medium. Hence the atom-atom interaction has a significant
accumulative effect only during the light storage, and one can neglect the
interaction effects during the light propagation. Under the EIT condition, the
induced coherence between the atomic ground and Rydberg $s$ state
$\rho_{sg}^{j}=-\Omega_{\mathrm{p}0}/\Omega_{\mathrm{c}}$ is proportional to
the Rabi frequency of the initial probe field $\Omega_{\mathrm{p}0}$
\cite{Lukin-03,Fleischhauer-05,Arimondo-PO-1996,Harris-PT-1997,Hau1999,Fleischhauer-PRL-2000,Fleischhauer2002}. 

During the propagation of the slow light pulse the probe and control beams
drive the $j$th atom to the dark state \cite{Fleischhauer2002}
\begin{equation}
|\mathrm{\Psi}^{j}\rangle=A\left(|g^{j}\rangle
-\frac{\Omega_{\mathrm{p}0}}{\Omega_{\mathrm{c}}}|s^{j}\rangle\right)
=A\left(1-\frac{\Omega_{\mathrm{p}0}}{\Omega_{\mathrm{c}}}
\sigma_{sg}^{j}\right)|g^{j}\rangle\,,
\label{eq:Dj}
\end{equation}
where
\begin{equation}
A=(1+\Omega_{\mathrm{p}0}^{2}/\Omega_{\mathrm{c}}^{2})^{-1/2}\label{eq:norm}
\end{equation}
is a normalization factor. Initially atoms are uncorrelated, thus the full
quantum state of the atomic ensemble is
\begin{equation}
|\Psi\rangle=\prod_{j}|\mathrm{\Psi}^{j}\rangle=A^{N}\prod_{j}
\left(1-\frac{\Omega_{\mathrm{p}0}}{\Omega_{\mathrm{c}}}\sigma_{sg}^{j}\right)|
\mathbf{g}\rangle\,,
\label{eq:state-all}
\end{equation}
where $N$ is a total number of atoms in the sample and
$|\mathbf{g}\rangle=\prod_{j}|g^{j}\rangle$ is a complete atomic ground state.
Note, that one can rewrite Eq.~(\ref{eq:state-all}) in terms of spin-wave
excitations:
\begin{equation}
|\Psi\rangle=A^{N}|\mathbf{g}\rangle+A^{N}\sum_{n=1}^{N}
\left(-\frac{\Omega_{\mathrm{p}0}}{\Omega_{c}}\right)^{n}|\Psi_{n}\rangle\,,
\end{equation}
where
\begin{equation}
|\Psi_{n}\rangle=\sum_{j_{1},\cdots,j_{n}}\sigma_{sg}^{j_{1}}\cdots\sigma_{sg}^{j_{n}}
|\mathbf{g}\rangle
\end{equation}
describes the state with $n$ spin excitations. Since the incident probe field
$\Omega_{\mathrm{p}}$ represents a pulse of a classical light, the full state
of the atoms $|\Psi\rangle$ is separable.

By switching off the control field, the probe field is stored in the atomic
coherences $\rho_{sg}^{j}$
\cite{Fleischhauer-PRL-2000,Hau-2001,Phillips-PRL-2001}. A quantum state of the
atomic ensemble is then given by Eq.~(\ref{eq:state-all}).  Immediately after
switching off the control laser, the $\pi/2$ optical pulse is applied that
couples the Rydberg $s$ and $p$ states and converts the Rydberg $s$ state to a
superposition of the $s$ and $p$ Rydberg states
$|+\rangle=(|s\rangle+|p\rangle)/\sqrt{2}$, as shown in
Fig.~\ref{fig:scheme}(b). Consequently, the state vector of the stored
dark-state polariton is converted to
\begin{equation}
|\Psi_{+}\rangle=A^{N}\prod_{j}\left(1-\frac{\Omega_{\mathrm{p}0}}{\Omega_{\mathrm{c}}}
\sigma_{+g}^{j}\right)
|\mathbf{g}\rangle\,.
\label{eq:state-all-after-first-pi/2}
\end{equation}
Here the symmetric an antisymmetric creation operators are defined
as
\begin{equation}
\sigma_{\pm g}^{j}=\frac{1}{\sqrt{2}}(\sigma_{sg}^{j}\pm\sigma_{pg}^{j})\,.
\label{eq:eq:sigma_g-pm}
\end{equation}
Note that the inverse transform reads
\begin{eqnarray}
\sigma_{sg}^{j} & = & \frac{1}{\sqrt{2}}(\sigma_{+g}^{j}+\sigma_{-g}^{j})\,,
\label{eq:sigma_gS}\\
\sigma_{pg}^{j} & = & \frac{1}{\sqrt{2}}(\sigma_{+g}^{j}-\sigma_{-g}^{j})\,.
\label{eq:sigma_gP}
\end{eqnarray}
Calling on Eq.~(\ref{eq:eq:sigma_g-pm}), it is convenient to represent
the initial state given by Eq.~(\ref{eq:state-all-after-first-pi/2})
in terms of the bare atomic states
\begin{equation}
|\Psi_{+}\rangle=A^{N}\prod_{j}\left(1-\frac{\Omega_{\mathrm{p}0}}{\sqrt{2}
\Omega_{\mathrm{c}}}(\sigma_{sg}^{j}+\sigma_{pg}^{j})\right)|\mathbf{g}\rangle\,.
\label{eq:state-all-after-first-pi/2-1}
\end{equation}

\subsection{Atomic state affected by the atom-atom interaction}

During the storage the atoms undergo a free evolution without influence of the
optical fields, yet affected by the atom-atom interaction
$\mathcal{H}_{\mathrm{at-at}}$ given by Eq.~(\ref{eq:ham-at-at}).  During such
an evolution the atomic state vector $|\Psi_{+}\rangle$ given by
Eq.~(\ref{eq:state-all-after-first-pi/2}) transforms to
\begin{equation}
|\Psi(T)\rangle=e^{-i\mathcal{H}_{\mathrm{at-at}}T}|\Psi_{+}\rangle\,,\label{eq:psi-T}
\end{equation}
where $T$ is the storage time. By collecting the terms containing double sums
as is detailed in Appendix~\ref{sec:appa} and using
Eq.~(\ref{eq:state-all-after-first-pi/2}), the action of the evolution operator
on the atomic state can be written as 
\begin{equation}
|\Psi(T)\rangle=|\Psi_{+}\rangle+\sum_{j\neq j'}
[e^{-iV(|\mathbf{r}_{j}-\mathbf{r}_{j'}|)T}-1]
\sigma_{ps}^{j}\sigma_{sp}^{j'}|\Psi_{+}\rangle+\mbox{nonpair terms.}
\label{eq:expansion-two-atoms-PRL}
\end{equation}
The terms that are not written explicitly in
Eq.~(\ref{eq:expansion-two-atoms-PRL}) contain triple and higher sums. In
Eq.~(\ref{eq:expansion-two-atoms-PRL}) we used the fact that the operators
$\sigma_{sg}^{j}$ and $\sigma_{pg}^{j}$ enter symmetrically the initial state
vector $|\Psi_{+}\rangle$ given by Eq.~(\ref{eq:state-all-after-first-pi/2-1}).
Consequently, the action of the operator $\sigma_{ps}^{j}\sigma_{sp}^{j'}$ on
the initial state vector $|\Psi_{+}\rangle$ gives the same result as the action
of the operator $\sigma_{pp}^{j}\sigma_{ss}^{j'}$ on $|\Psi_{+}\rangle$.  This
allows us to combine $\cos[V(|\mathbf{r}_{j}-\mathbf{r}_{j'}|)T]$ and
$-i\sin[V(|\mathbf{r}_{j}-\mathbf{r}_{j'}|)T]$ entering
Eq.~(\ref{eq:expansion-two-atoms}) into a single exponential function when
acting on the initial state vector $|\Psi_{+}\rangle$.

From now on we will omit the terms due to the interaction involving three and
more atoms.  Such an assumption is legitimate if the density of Rydberg atoms
is small enough and the duration of the storage time $T$ is sufficiently short,
so that it is unlikely to have more than a single pair of strongly interacting
closeby Rydberg atoms. This is the case if a characteristic distance
$r_{\mathrm{c}}$, at which the RDDI potential $V(r_{\mathrm{c}})$ becomes
equal to the inverse storage time $T^{-1}$, is smaller than a mean distance
$r_{\mathrm{Ry}}=n_{\mathrm{Ry}}^{-1/3}$ between the atoms excited to the
Rydberg state, i.e. $r_{\mathrm{c}}\lesssim r_{\mathrm{Ry}}$.  Here
\begin{equation}
n_{\mathrm{Ry}}=A^{2}\frac{\Omega_{\mathrm{p}0}^{2}}{\Omega_{\mathrm{c}}^{2}}n
\label{eq:rydberg-density}
\end{equation}
is a density of Rydberg atoms, $n$ is a total density of atoms, and
$\Omega_{\mathrm{p}0}^{2}/\Omega_{\mathrm{c}}^{2}\ll1$ is a probability for an
individual atom to be excited to the Rydberg $s$ state during the initial
propagation of slow light. The RDDI potential depends on the distance between
atoms as $V(r)=C_{3}/r^{3}$, giving a characteristic distance
\begin{equation}
r_{\mathrm{c}}=(C_{3}T)^{\frac{1}{3}}\,.
\end{equation}
The condition $r_{\mathrm{c}}\lesssim r_{\mathrm{Ry}}$ then leads to an upper
limit for the storage time $T_{\mathrm{max}}=(C_{3}n_{\mathrm{Ry}})^{-1}$.
A more detailed discussion of the validity is given in Appendix~\ref{sec:appb}.

In writing Eq.~(\ref{eq:expansion-two-atoms-PRL}) we have included multiple
RDDI transitions within the same pair of atoms providing an oscillating
term $\exp[-iV(|\mathbf{r}_{j}-\mathbf{r}_{j'}|)T]$ which regularizes
a divergent behavior of individual terms in the expansion of the evolution
operator in the powers of $T$. This is important at interatomic distances
$|\mathbf{r}_{j}-\mathbf{r}_{j'}|<r_{\mathrm{c}}$ for which the RDDI
energy $V(|\mathbf{r}_{j}-\mathbf{r}_{j'}|)$ exceeds $T^{-1}$.

\begin{figure}
\includegraphics[width=0.6\textwidth]{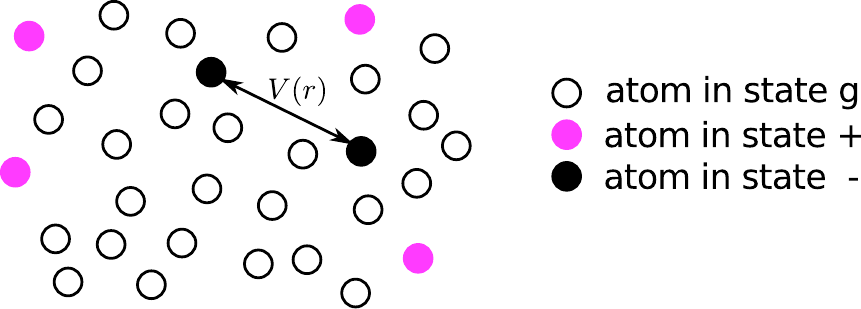}
\caption{Schematic depiction of the state of atoms at the end of storage period.}
\label{fig:state}
\end{figure}

Taking into account Eq.~(\ref{eq:state-all-after-first-pi/2-1}) for the initial
state vector $|\Psi_{+}\rangle$, Eq.~(\ref{eq:expansion-two-atoms-PRL}) yields
the following state vector of the atomic system at the end of the free
evolution:
\begin{equation}
|\Psi(T)\rangle\approx|\Psi_{+}\rangle+A^{N}
\frac{\Omega_{\mathrm{p}0}^{2}}{2\Omega_{\mathrm{c}}^{2}}
\sum_{j\neq j'}[e^{-iV(|\mathbf{r}_{j}-\mathbf{r}_{j'}|)T}-1]\sigma_{sg}^{j}\sigma_{pg}^{j'}
\prod_{j''\neq j,j'}\left(1-\frac{\Omega_{\mathrm{p}0}}{\Omega_{\mathrm{c}}}
\sigma_{+g}^{j''}\right)|\mathbf{g}\rangle\,.
\label{eq:state-final-0}
\end{equation}
Using Eqs.~(\ref{eq:sigma_gS}) and (\ref{eq:sigma_gP}), a pair of operators
$\sigma_{sg}^{j}\sigma_{pg}^{j'}$ entering the above equation can be cast in
terms of the operators $\sigma_{\pm g}^{j}$ and $\sigma_{\pm g}^{j^{\prime}}$
as
\begin{equation}
\sigma_{sg}^{j}\sigma_{pg}^{j'}=(\sigma_{+g}^{j}+\sigma_{-g}^{j})(\sigma_{+g}^{j'}
-\sigma_{-g}^{j'})/2\,.
\label{eq:pair-of-operators}
\end{equation}
Since the summation indices can be interchanged, the mixed terms containing the
operators $-\sigma_{+g}^{j}\sigma_{-g}^{j'}$ and
$\sigma_{-g}^{j}\sigma_{+g}^{j'}$ cancel each other in
Eq.~(\ref{eq:state-final-0}), so $\sigma_{sg}^{j}\sigma_{pg}^{j'}$ can be
replaced by
$\left(\sigma_{+g}^{j}\sigma_{+g}^{j'}-\sigma_{-g}^{j}\sigma_{-g}^{j'}\right)/2$,
giving
\begin{equation}
|\Psi(T)\rangle\approx|\Psi_{+}\rangle+A^{N}
\frac{\Omega_{\mathrm{p}0}^{2}}{4\Omega_{\mathrm{c}}^{2}}
\sum_{j\neq j'}[e^{-iV(|\mathbf{r}_{j}-\mathbf{r}_{j'}|)T}-1](\sigma_{+g}^{j}
\sigma_{+g}^{j'}-\sigma_{-g}^{j}\sigma_{-g}^{j'})\prod_{j''\neq j,j'}
\left(1-\frac{\Omega_{\mathrm{p}0}}{\Omega_{\mathrm{c}}}\sigma_{+g}^{j''}\right)
|\mathbf{g}\rangle\,.
\label{eq:state-final-2}
\end{equation}
The states of the atoms at the end of the free evolution are schematically
shown in Fig.~\ref{fig:state}.

Just before the retrieval one applies another $\pi/2$ optical pulse that
couples the Rydberg states $s$ and $p$. This converts the state
$|+\rangle=(|s\rangle+|p\rangle)/\sqrt{2}$ into the Rydberg state $|p\rangle$,
whereas the state $|-\rangle=(|s\rangle-|p\rangle)/\sqrt{2}$ is converted back
into the Rydberg state $|s\rangle$ (see
Fig.~\ref{fig:scheme}(b)).  As a result the state vector
(\ref{eq:state-final-2}) reduces to
\begin{equation}
|\Psi_{\mathrm{fin}}\rangle\approx|\Psi_{p}\rangle+A^{N}
\frac{\Omega_{\mathrm{p}0}^{2}}{4\Omega_{\mathrm{c}}^{2}}
\sum_{j\neq j'}[e^{-iV(|\mathbf{r}_{j}-\mathbf{r}_{j'}|)T}-1](\sigma_{pg}^{j}\sigma_{pg}^{j'}
-\sigma_{sg}^{j}\sigma_{sg}^{j'})\prod_{j''\neq j,j'}
\left(1-\frac{\Omega_{\mathrm{p}0}}{\Omega_{\mathrm{c}}}\sigma_{pg}^{j''}\right)
|\mathbf{g}\rangle
\label{eq:state-final-converted}
\end{equation}
where $|\Psi_{p}\rangle$ is obtained from
Eq.~(\ref{eq:state-all-after-first-pi/2}) with $\sigma_{+g}^{j}$ replaced by
$\sigma_{pg}^{j}$:
\begin{equation}
|\Psi_{p}\rangle=A^{N}\prod_{j}\left(1-\frac{\Omega_{\mathrm{p}0}}{\Omega_{\mathrm{c}}}
  \sigma_{pg}^{j}\right)|\mathbf{g}\rangle\,.
\end{equation}
The second term in Eq.~(\ref{eq:state-final-converted})
represents the correlated pairs of atoms in the Rydberg $s$ and $p$ states
created due to the atom-atom interaction. As we shall see in
Sec.~\ref{sec:retrieval}, the Rydberg $s$ excitations are converted into pairs
of correlated photons during the restoring of the slow light. The first term in
Eq.~(\ref{eq:state-final-converted}) does not contain the Rydberg $s$ state and
thus will not contribute to the restored slow light.

\subsection{Atomic correlation functions}

The spectral density of the restored light is related to the atomic
first-order correlation at different sites $j$ and $j^{\prime}$,
\begin{equation}
G_{\mathrm{at}}^{(1)}(\mathbf{r}_{j},\mathbf{r}_{j'})=
\langle\Psi_{\mathrm{fin}}|\sigma_{gs}^{j\dag}\sigma_{gs}^{j'}|
\Psi_{\mathrm{fin}}\rangle\,.
\label{eq:atom-first-order}
\end{equation}
Using the approximate expression~(\ref{eq:state-final-converted})
for the final state we get 
\begin{equation}
G_{\mathrm{at}}^{(1)}(\mathbf{r}_{j},\mathbf{r}_{j'})\approx A^{4}
\frac{\Omega_{\mathrm{p}0}^{4}}{4\Omega_{\mathrm{c}}^{4}}
\sum_{j''\neq j,j'}[e^{iV(|\mathbf{r}_{j}-\mathbf{r}_{j''}|)T}-1]
[e^{-iV(|\mathbf{r}_{j'}-\mathbf{r}_{j''}|)T}-1]\,.
\label{eq:atom-first-order-2}
\end{equation}
The intensity of the restored light is related to the density of atoms
in the Rydberg state $|s\rangle$. The probability for an atom to
be in this state can be obtained from the atomic correlation function
at the same site $j$:
\begin{equation}
G_{\mathrm{at}}^{(1)}(\mathbf{r}_{j})\equiv
G_{\mathrm{at}}^{(1)}(\mathbf{r}_{j},\mathbf{r}_{j})=
\langle\Psi_{\mathrm{fin}}|\sigma_{gs}^{j\dag}\sigma_{gs}^{j}|\Psi_{\mathrm{fin}}\rangle\,.
\end{equation}
Using Eq.~(\ref{eq:atom-first-order-2}) this probability reduces
to 
\begin{equation}
G_{\mathrm{at}}^{(1)}(\mathbf{r}_{j})\approx
A^{4}\frac{\Omega_{\mathrm{p}0}^{4}}{2\Omega_{\mathrm{c}}^{4}}\sum_{j'\neq j}
\left(1-\cos[V(|\mathbf{r}_{j}-\mathbf{r}_{j'}|)T]\right)\,.
\label{eq:atom-prob-2}
\end{equation}
The atomic second-order correlation function
\begin{equation}
G_{\mathrm{at}}^{(2)}(\mathbf{r}_{j},\mathbf{r}_{j'})=
\langle\Psi_{\mathrm{fin}}|\sigma_{gs}^{j\dag}\sigma_{gs}^{j'\dag}
\sigma_{gs}^{j'}\sigma_{gs}^{j}|\Psi_{\mathrm{fin}}\rangle
\end{equation}
using Eq.~(\ref{eq:state-final-converted}) for
$|\Psi_{\mathrm{fin}}\rangle$ can be expressed as
\begin{equation}
G_{\mathrm{at}}^{(2)}(\mathbf{r}_{j},\mathbf{r}_{j'})\approx(1-\delta_{j,j'})A^{4}
\frac{\Omega_{\mathrm{p}0}^{4}}{2\Omega_{\mathrm{c}}^{4}}
\left(1-\cos[V(|\mathbf{r}_{j}-\mathbf{r}_{j'}|)T]\right)\,.
\label{eq:atom-corr-1}
\end{equation}
The atomic second-order correlation function
(\ref{eq:atom-corr-1}) is related to the second-order correlation function of
the restored light $G^{(2)}(\tau)$ via Eq.~(\ref{eq:g2-average}) presented
below.

\section{Properties of restored probe pulse}
\label{sec:retrieval}

After application of the second  $\pi/2$  pulse the probe pulse of light is
restored by switching on the control beam characterized by the Rabi frequency
$\Omega_{\mathrm{c}}$.  The state of the atomic ensemble just before
the retrieval is given by Eq.~(\ref{eq:state-final-converted}). During the
retrieval only the atoms in the Rydberg state $s$ contribute to the probe beam.
The $p$ state excitations remain in the medium and thus will no longer be
considered
\footnote{During propagation of the regenerated beam there is some probability
for the RDDI to exchange the $s$ Rydberg state with $p$ Rydberg states present
in the medium. Yet the RDDI can alter the location of the $s$ state by a radius
much smaller than the distance $v_{g0}\tau$ which is spectroscopically probed
by measuring the second order correlation function of the regenerated light.
Therefore the RDDI between the $s$ and $p$ Rydberg states should not yield a
noticeable affect on the second order correlation function of the regenerated
beam.}. The restored probe field
\begin{equation}
\Omega_{\mathrm{p}}(\mathbf{\mathbf{r}}_{j})=-\Omega_{\mathrm{c}}\sigma_{gs}^{j}(T)
\end{equation}
is generated from atomic coherences $\sigma_{gs}^{j}(T)$ involving the Rydberg
$s$ state.

Note that the light restored from the atomic state
(\ref{eq:state-final-converted}) consist of pairs of correlated photons
corresponding to the second term in Eq.~(\ref{eq:state-final-converted}), there
being no contributions by single photons. Single photons can appear only due to
losses, when one of the photons forming the pair is absorbed. Yet, as we will
see later in this section, the absorption does not significantly distort the
second-order correlation function for a sufficiently large separation between
the photons.

The second term of Eq.~(\ref{eq:state-final-converted}) describes a
superposition of all possible atomic pairs, so the restored light is
in a quantum superposition of all corresponding photon pairs. The measurement
of the second-order correlation function of the restored light selects a photon pair
in which the photons are separated by a chosen distance.

\subsection{Second-order correlation function of the restored light}

\begin{figure}
\includegraphics[width=0.4\textwidth]{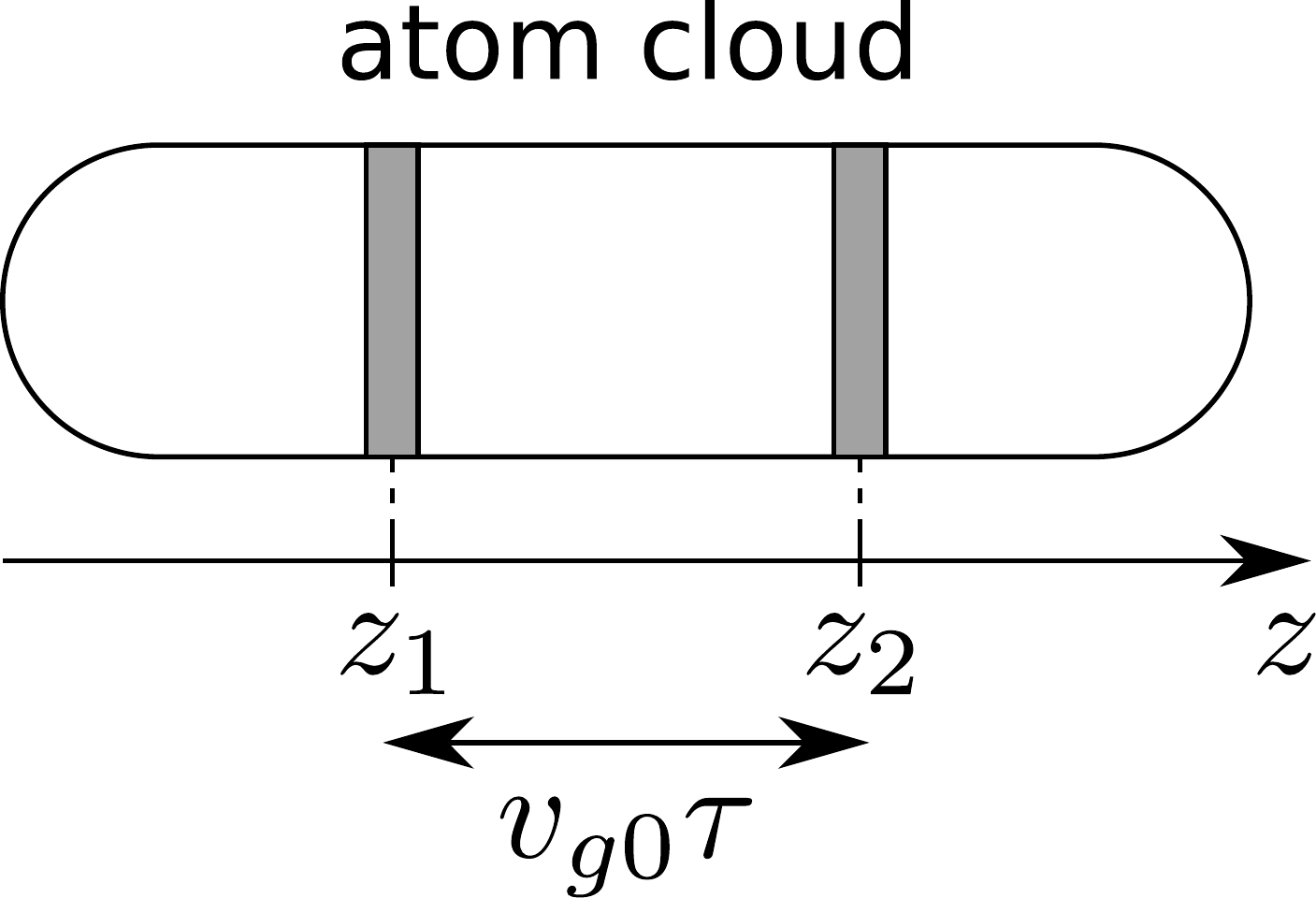}
\caption{Regions (gray color) over which the atomic second-order correlation
function $G_{\mathrm{at}}^{(2)}$ is averaged in calculating the second-order
correlation function of the restored light.}
\label{fig:region}
\end{figure}

The second-order correlation function of the retrieved light
\begin{equation}
G^{(2)}(\tau)=\langle\Omega_{\mathrm{p}}^{\dag}(t)\Omega_{\mathrm{p}}^{\dag}(t+\tau)
\Omega_{\mathrm{p}}(t+\tau)\Omega_{\mathrm{p}}(t)\rangle
\end{equation}
is calculated by averaging the atomic second-order correlation function
$G_{\mathrm{at}}^{(2)}(\mathbf{r}_{j},\mathbf{r}_{j'})=\langle\Psi_{\mathrm{fin}}|
\sigma_{gs}^{j\dag}\sigma_{gs}^{j'\dag}\sigma_{gs}^{j'}\sigma_{gs}^{j}
|\Psi_{\mathrm{fin}}\rangle$ over the atomic positions $\mathbf{r}_{j}$ and
$\mathbf{r}_{j'}$ separated by $|z_{j}-z_{j'}|=v_{g0}\tau$ along the
propagation direction $z$. Thus the correlation function of the restored light
reads
\begin{equation}
G^{(2)}(\tau)=\Omega_{\mathrm{c}}^{4}/N_{r}
\sum_{j,j'}G_{\mathrm{at}}^{(2)}(\mathbf{r}_{j},\mathbf{r}_{j'})\,,
\label{eq:g2-average}
\end{equation}
where the summation extends over a narrow region, shown in
Fig.~\ref{fig:region}, in which the atomic second-order correlation function is
averaged. Here $N_{r}$ is the number of atoms used in averaging.  If the width
of the atomic medium is smaller than the separation distance $v_{g0}\tau$, the
averaging does not significantly alter the atomic second-order correlation
function $G_{\mathrm{at}}^{(2)}(\mathbf{r}_{j},\mathbf{r}_{j'})$. By
concentrating on the distances between the atoms $|\mathbf{r}-\mathbf{r}'|$
larger than the width of the atom cloud the problem becomes essentially
one-dimensional. 

From Eq.~(\ref{eq:atom-corr-1}) follows that  the second-order correlation
function of the retrieved light is determined by the atom-atom interaction:
\begin{equation}
G^{(2)}(\tau)\propto1-\cos[V(v_{g0}\tau)T]\,.
\end{equation}
One can see that the scale of the distances $v_{g0}\tau$ probed by the
second-order correlation function depends on the storage time $T$.  For a
sufficiently small storing time $T$ and a large delay time $\tau$,
$V(v_{g0}\tau)T\ll\pi$, the second-order correlation function of the retrieved
light is proportional to the square of the interaction potential at the
interatomic distance $v_{g0}\tau$,  i.e.,
$G^{(2)}(\tau)\propto[V(v_{g0}\tau)T]^{2}$.  In this way the restored pulse is
created exclusively due to the atom-atom interaction which vanishes as
interatomic distance increases. The probability to find in the restored pulse
a pair of photons separated by large distances goes to zero, and the
(unnormalized) second-order correlation function $G^{(2)}(\tau)$ decays as
$\tau$ increases.

\subsection{Estimation of the intensity of the restored light}

Let us estimate the intensity of the restored probe pulse. The restored field
is generated from atomic coherences $\sigma_{gs}^{j}(T)$ involving the Rydberg
$s$ state, according to the equation
$\Omega_{\mathrm{p}}(\mathbf{\mathbf{r}}_{j})=-\Omega_{\mathrm{c}}\sigma_{gs}^{j}(T)$.
Thus the intensity, proportional to $|\Omega_{\mathrm{p}}|^2$, can be
calculated using atomic first order correlation function
$G_{\mathrm{at}}^{(1)}$. Calling on Eq.~(\ref{eq:atom-prob-2}) the ratio of the
intensities of the restored and the incoming probe pulses is given by
\begin{equation}
\frac{\Omega_{\mathrm{p},\mathrm{out}}^{2}}{\Omega_{\mathrm{p}0}^{2}}=A^4
\frac{\Omega_{\mathrm{p}0}^{2}}{2\Omega_{\mathrm{c}}^{2}}
\sum_{j}\left(1-\cos[V(|\mathbf{r}-\mathbf{r}_{j}|)T]\right)\,.
\label{eq:intensity-restored}
\end{equation}
The sum in Eq.~(\ref{eq:intensity-restored}) can be estimated as
\begin{equation}
\sum_{j}\left(1-\cos[V(|\mathbf{r}-\mathbf{r}_{j}|)T]\right)
\approx n
\int d\mathbf{r}'\left(1-\cos[V(|\mathbf{r}-\mathbf{r}'|)T]\right)
\sim n r_{\mathrm{c}}^3\,,
\end{equation}
where we have used Eq.~(\ref{eq:integral-1}) for evaluating the integral.
Using Eq.~(\ref{eq:rydberg-density}) for the density of Rydberg atoms, we
obtain that the ratio of the intensities of the restored and the incoming probe
pulses is of the order of
\begin{equation}
\frac{\Omega_{\mathrm{p},\mathrm{out}}^{2}}{\Omega_{\mathrm{p}0}^{2}}\sim
n_{\mathrm{Ry}}r_{\mathrm{c}}^3\,.
\end{equation}
Note, that in order to neglect the interaction involving three or more Rydberg
atoms we require that $n_{\mathrm{Ry}}r_{\mathrm{c}}^3\ll 1$. Thus the
intensity of the restored light is much smaller than that of the incident
light. Consequently most of the probe pulse remains in the medium in the form
of excitations of the Rydberg $p$ state.

\subsection{Spectral width of the restored light}
\label{sub:spectum}

The spectrum $S(\omega)$ is related to the first-order correlation function of
the light
\begin{equation}
G^{(1)}(\tau)=\langle\Omega_{\mathrm{p}}^{\dag}(t+\tau)
\Omega_{\mathrm{p}}(t)\rangle
\end{equation}
via the equation
\begin{equation}
S(\omega)=\int e^{-i\omega\tau}G^{(1)}(\tau)d\tau\,.
\end{equation}
The restored probe field
$\Omega_{\mathrm{p}}(\mathbf{\mathbf{r}}_{j})=-\Omega_{\mathrm{c}}\sigma_{gs}^{j}(T)$
is generated from atomic coherences $\sigma_{gs}^{j}(T)$ involving the Rydberg
$s$ state.  Therefore the spectral width can be calculated from the atomic
first order correlation function $G_{\mathrm{at}}^{(1)}$ at different sites.
Using Eq.~(\ref{eq:atom-first-order-2}) we get
\begin{equation}
G^{(1)}(\tau)\sim\sum_{j}[e^{iV(|\mathbf{r}-\mathbf{r}_{j}|)T}-1]
[e^{-iV(|\mathbf{r}-\mathbf{r}_{j}+v_{g0}\tau\hat{\mathbf{e}}_{z}|)T}-1]\,.
\label{eq:g1-restored}
\end{equation}
We see that the restored light acquires a finite width of the spectrum, even
when the incident probe beam is monochromatic. From Eq.~(\ref{eq:g1-restored})
one can estimate the characteristic width of the function $G^{(1)}(\tau)$ to be
of the order of $r_{\mathrm{c}}/v_{g0}$, where $r_{\mathrm{c}}$ is defined as
$V(r)T=(r_{\mathrm{c}}/r)^{3}$. Thus the spectral width of the restored light
$S(\omega)$ is of the order of $v_{g0}/r_{\mathrm{c}}$.

\subsection{Influence of losses for the two photon correlation measurements}

Since the restored light has a finite spectral width $v_{g0}/r_{\mathrm{c}}$,
it experiences losses due to a finite transmittivity width of the EIT window.
To estimate the influence of losses occurring during the propagation of the
restored light in the atomic medium, let us consider a part of the atomic state
containing only pairs of atoms in the Rydberg $s$ state.  According to
Eq.~(\ref{eq:state-final-converted}), this part is
\begin{equation}
|\Psi_{\mathrm{ss}}\rangle=\frac{\Omega_{\mathrm{p}0}^{2}}{4\Omega_{\mathrm{c}}^{2}}
\sum_{j\neq j'}[e^{-iV(|\mathbf{r}_{j}-\mathbf{r}_{j'}|)T}-1]\sigma_{gs}^{j'\dag}
\sigma_{gs}^{j\dag}|\mathbf{g}\rangle\,.
\label{eq:psi-UU}
\end{equation}
We are interested in the delay times $\tau$ entering the second-order photon
correlation function
$G^{(2)}(\tau)=\langle\Omega_{\mathrm{p}}^{\dag}(t)\Omega_{\mathrm{p}}^{\dag}(t+\tau)
\Omega_{\mathrm{p}}(t+\tau)\Omega_{\mathrm{p}}(t)\rangle$, such that
$v_{g0}\tau$ is larger than the width of the atom cloud.  Such delay times
correspond to the distances between atoms $|\mathbf{r}-\mathbf{r}'|$ larger
than the width of the medium. In this case we can consider the state
$|\Psi_{\mathrm{ss}}\rangle$ written in terms of the spin-flip operators
averaged over the cross section of the medium:
\begin{equation}
|\Psi_{ss}\rangle=C\int dz\int dz'I(|z-z'|)
\sigma_{gs}^{\dag}(z')\sigma_{gs}^{\dag}(z)|\mathbf{g}\rangle\,,
\label{eq:psi-two}
\end{equation}
where
\begin{equation}
I(z)\equiv i\left(e^{-iV(z)T}-1\right)\,,\qquad C=
\frac{\Omega_{\mathrm{p}0}^{2}}{4\Omega_{\mathrm{c}}^{2}}n^{2}S^{2}\,.
\end{equation}
Introducing the spin-flip operators in the momentum representation
\begin{equation}
\sigma_{gs,k}=\frac{1}{\sqrt{2\pi}}\int dz\, e^{ikz}\sigma_{gs}(z)
\end{equation}
and using Eq.~(\ref{eq:psi-two}) we get
\begin{equation}
|\Psi_{ss}\rangle=\frac{C}{2\pi}\int dk\int dk'\int dz\int dz'
I(|z-z'|)e^{ikz+ik'z'}\sigma_{gs,k}^{\dag}\sigma_{gs,k'}^{\dag}|\mathbf{g}\rangle\,.
\end{equation}
By separating the mean momentum $\bar{k}=\frac{1}{2}(k+k')$ and the difference
$\Delta k=\frac{1}{2}(k-k')$, the atomic state $|\Psi_{ss}\rangle$ can be
written as
\begin{equation}
|\Psi_{ss}\rangle=\frac{C}{2\pi}\int d\Delta k\,\tilde{I}(\Delta k)
\sigma_{gs,\Delta k}^{\dag}\sigma_{gs,-\Delta k}^{\dag}|\mathbf{g}\rangle\,,
\end{equation}
where
\begin{equation}
\tilde{I}(k)=i\int dz\,[e^{-iV(|z|)T}-1]e^{ikz}
\end{equation}
is the Fourier transform of $I(z)$. After restoring the light, the spin-flip
operators $\sigma_{gs,k}$ are replaced by operators for dark-state polaritons
$\mathcal{P}_{\mathrm{D},k}=\frac{v_{g0}}{c}
\frac{\Omega_{\mathrm{p},k}}{\Omega_{\mathrm{c}}}-\sigma_{gs,k}$, giving the
state 
\begin{equation}
|\Phi\rangle=\frac{C}{2\pi}\int dk\,\tilde{I}(k)\mathcal{P}_{\mathrm{D},k}^{\dag}
\mathcal{P}_{\mathrm{D},-k}^{\dag}|\mathrm{vac}\rangle\,.
\end{equation}
This state describes a pair of polaritons with the wave vectors $k$ and $-k$
around the zero central wave vector: $\bar{k}=0$.

The propagation duration of the restored polariton is of the order of
\begin{equation}
\tau_{\mathrm{prop}}\sim\frac{L}{2v_{g0}}
\end{equation}
assuming that it propagates approximately half of a medium length $L$. During
the propagation the dark-state polaritons decay due to nonadiabatic losses
with the rate
$\gamma_{\mathrm{pol}}=2\Gamma(v_{g0}k)^{2}/\Omega_{\mathrm{c}}^{2}$
\cite{Juzeliunas2002}. Thus the polariton operator $\mathcal{P}_{\mathrm{D},k}$
changes to 
\begin{equation}
\mathcal{P}_{\mathrm{D},k}e^{-\frac{L^{2}}{2\alpha}k^{2}}
+\eta_{k}b_{k}\,,\quad\mbox{where}\quad\eta_{k}
=\sqrt{1-e^{-\frac{L^{2}}{\alpha}k^{2}}}\,.
\end{equation}
The exponent $L^{2}k^{2}/(2\alpha)=\gamma_{\mathrm{pol}}\tau_{\mathrm{prop}}$
describes the polariton decay. Additional bosonic noise operators $b_{k}$ have
been included to preserve the commutation relations \cite{Scully1997}. The
final state then becomes
\begin{equation}
|\Phi'\rangle\sim\int dk\,\tilde{I}(k)\left(\mathcal{P}_{\mathrm{D},k}^{\dag}
  e^{-\frac{L^{2}}{2\alpha}k^{2}}
  +\eta_{k}b_{k}^{\dag}\right)\left(\mathcal{P}_{\mathrm{D},-k}^{\dag}
  e^{-\frac{L^{2}}{2\alpha}k^{2}}
  +\eta_{k}b_{-k}^{\dag}\right)|\mathrm{vac}\rangle\,.
\end{equation}
We are interested in the second-order correlation function
\begin{equation}
G^{(2)}(z,z')=\langle\Phi'|\mathcal{P}_{\mathrm{D}}^{\dag}(z)
\mathcal{P}_{\mathrm{D}}^{\dag}(z')
\mathcal{P}_{\mathrm{D}}(z')\mathcal{P}_{\mathrm{D}}(z)|\Phi'\rangle\,.
\label{eq:g2-supp}
\end{equation}
It is noteworthy that only the component of the state vector $|\Phi'\rangle$
containing two polariton operators,
\begin{equation}
\int dk\,\tilde{I}(k)\mathcal{P}_{\mathrm{D},k}^{\dag}\mathcal{P}_{\mathrm{D},-k}^{\dag}
e^{-\frac{L^{2}}{\alpha}k^{2}}|\mathrm{vac}\rangle\,,
\end{equation}
contributes to $G^{(2)}(z,z')$. Therefore, going back to the coordinate
representation we obtain
\begin{equation}
\int dz\int dz'\, I'(z-z')\mathcal{P}_{\mathrm{D}}^{\dag}(z)
\mathcal{P}_{\mathrm{D}}^{\dag}(z')|\mathrm{vac}\rangle\,,
\label{eq:polariton-state-with-losses}
\end{equation}
where
\begin{equation}
I'(z)\approx\frac{1}{2\pi}\int dk\,\tilde{I}(k)e^{-ikz-\frac{L^{2}}{\alpha}k^{2}}
=i\frac{\sqrt{\alpha}}{2L\sqrt{\pi}}\int d\tilde{z}\,[e^{-iV(|\tilde{z}|)T}-1]
e^{-\frac{\alpha}{4L^{2}}(\tilde{z}-z)^{2}}\,.
\label{eq:potent-modif}
\end{equation}
We see that losses modify the coefficient $e^{-iV(z)T}-1$ appearing in
Eq.~(\ref{eq:state-final-converted}) by convolving it with a Gaussian
$\exp[-\alpha(z/2L)^{2}]$.  Combining Eqs.~(\ref{eq:g2-supp}) and
(\ref{eq:polariton-state-with-losses}), we obtain that the polariton losses
change the second-order correlation function $|I(z-z')|^{2}$ to
$G^{(2)}(z,z')=|I'(z-z')|^{2}$.  Consequently the second-order correlation
function of the restored light is determined by a modified potential influenced
by the losses, rather than by an actual interaction potential $V$.

\begin{figure}
\includegraphics[width=0.6\textwidth]{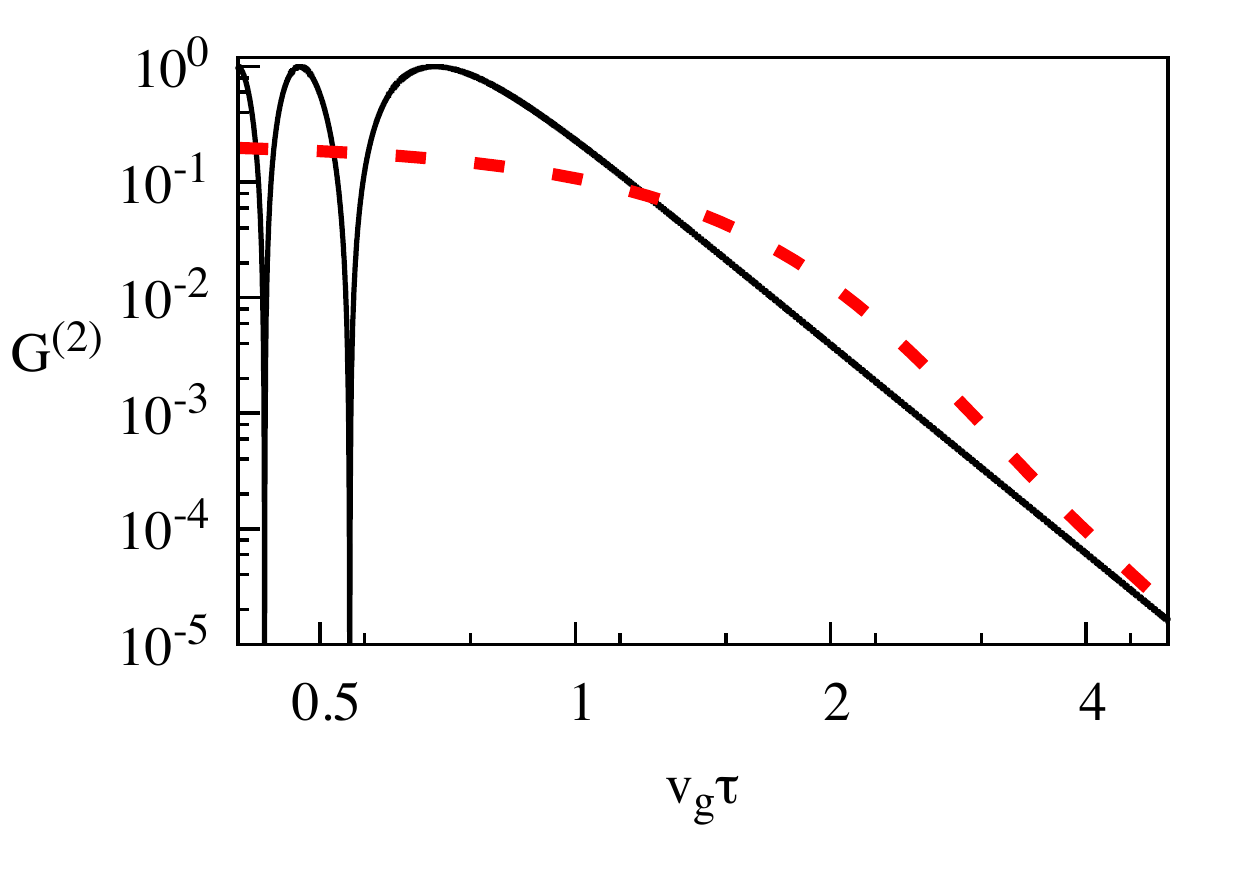}
\caption{(color online). The second-order correlation function of the restored
light normalized to the intensity of the input pulse without including
losses (solid black line) and modified by polariton losses (dashed
red line). The distances are measured in units of $r_{\mathrm{c}}$,
the characteristic distance of losses is assumed to be $L/\sqrt{\alpha}=1/2$.}
\label{fig:modif-potent}
\end{figure}

The second-order correlation function with and without inclusion of losses is
displayed in Fig.~\ref{fig:modif-potent}.
In order to get a dimensionless quantity, the second-order
correlation function shown in Fig.~\ref{fig:modif-potent} is normalized using
the intensity of the initial probe pulse.
As the dashed red curve in
Fig.~\ref{fig:modif-potent} indicates, the losses during the propagation of
light lead to the smoothening of the second-order correlation function, in
agreement with Eq.~(\ref{eq:potent-modif}).
One can see that the the influence of losses diminishes for distances
$z=v_{g0}\tau$ much larger than a characteristic loss distance
$L/\sqrt{\alpha}$.  This corresponds to a delay time $\tau$ greater than
$\Gamma\sqrt{\alpha}/\Omega_{\mathrm{c}}^{2}$.  For smaller delay times $\tau$
the effects of the losses become significant, reducing the number of photon
pairs.

Note that the spectral width of the restored light, considered
in the Sec.~\ref{sub:spectum}, is related to the first-order correlation
function rather than to the second-order correlation function shown in
Fig.~\ref{fig:modif-potent}. In general those two correlation functions are not
directly related; only for chaotic classical light sources is the second-order
correlation $g^{(2)}$  determined by the first-order correlation $g^{(1)}$
\cite{Loudon2000,Scully1997}.
Thus, in the absence of losses the spectral width does not limit the structure
of the second-order correlation function. This is not the case when losses are
present, as one can see from the red dashed curve in
Fig.~\ref{fig:modif-potent}.

\section{Discussion and conclusions}
\label{sec:concl}

The proposed ladder scheme can be experimentally implemented using ultracold
$^{87}\mathrm{Rb}$ atoms \cite{Peyronel2012} by preparing the atoms in a
hyperfine ground state $|5S_{1/2},F=2\rangle$ serving as the state $|g\rangle$
in our scheme. A hyperfine excited state $|5P_{3/2},F=3\rangle$ with a decay
rate $\Gamma=2\pi\times6\,\mathrm{MHz}$ corresponds to the state $|e\rangle$.
The characteristic distance $r_{\mathrm{c}}$ of the order
$r_{\mathrm{c}}\approx0.18\,\mathrm{mm}$ can be achieved for the storage
duration $T$ of the order of $10\,\mu\mathrm{s}$ and a principal quantum number
of the Rydberg levels $n=100$, the latter leading to the coefficient $C_{3}$ of
the order of $610\,\mathrm{GHz}\cdot\mu\mathrm{m}^{3}$.  The interaction
potential $V(r_{\mathrm{c}})=0.1\,\mathrm{MHz}$ can be much smaller than the
microwave Rabi frequency, which has been created of the order of
$100\,\mathrm{MHz}$ \cite{Sedlacek2012,Maxwell2013}.  Note, that for such a
large principal quantum number a strong Rydberg blockade occurs, with the
blockade radius of the order of $13\,\mu\mathrm{m}$ \cite{Peyronel2012}. Yet
here we are spectroscopically probing the inter-atomic distances larger than
the Rydberg blockade radius, so the blockade effects are not important. On the
other hand, the polariton losses due to the finite spectral width of the
regenerated light can be neglected for distances between the emitting atomic
pairs $v_{g0}\tau$ larger than $0.18\,\mathrm{mm}$ when using the
experimentally accessible length of the atomic medium $L=1\,\mathrm{mm}$ and
the optical density $\alpha=30$ \cite{Chiu2014,Hsiao2014}. The Rabi frequency
of the control beam $\Omega_{\mathrm{c}}=2\pi\times2\,\mathrm{MHz}$ leads to
the group velocity of the polaritons $v_{g0}=140\,\mathrm{m}/\mathrm{s}$, thus
these distances correspond to the delay time $\tau\approx1.3\,\mu\mathrm{s}$.
In this way, it is feasible to observe correlated photon pairs produced by
storing and regenerating the Rydberg slow light.

The proposed method can be applied not only to the resonant dipole-dipole
interactions but also to the other types of atom-atom interactions.  The
suggested Ramsey-type scheme can be employed to generate narrow-linewdth
biphotons with correlation times of the order of the propagation delay time.
The scheme can also provide an efficient way for manipulation of individual
photons or operation of qubits, since two photons can interact with each other
effectively at relatively large distances determined by the interaction between
the Rydberg atoms. With increasing the storage time (e.g.\ by using an optical
lattice to confine the atomic motion within a distance smaller than the
wavelength \cite{Schauss2012}), the scheme may be used as a sensitive tool for
probing the atom-atom interaction.  In this way our proposal offers new
possibilities and novel applications in generation of nonclassical light,
manipulation of quantum information, and precision measurement of long-distance
interaction.

\begin{acknowledgments}
Helpful discussions with Trey Porto are gratefully acknowledged.  Authors also
acknowledge many fruitful discussions on this work under the platforms of TG7
and E1 programs sponsored by National Center for Theoretical Sciences, Taiwan.
This work was supported by Project No.\ TAP LLT-2/2016 of the Research Council
of Lithuania and the Ministry of Science and Technology of Taiwan under Grants
No.\ 104-2119-M-007-004 and No.\ 105-2923-M-007-002-MY3.
\end{acknowledgments}

\appendix

\section{Interaction of a pair of atoms}
\label{sec:appa}

Let us consider the free evolution operator
$e^{-i\mathcal{H}_{\mathrm{at-at}}T}$.  For further approximations it is
convenient to represent the operator $e^{-i\mathcal{H}_{\mathrm{at-at}}T}$ as
\begin{equation}
e^{-i\mathcal{H}_{\mathrm{at-at}}T}=1+(e^{-i\mathcal{H}_{\mathrm{at-at}}T}-1)\,.
\label{eq:evolution-free-manipulation}
\end{equation}
Since the duration of the evolution $T$ is considered to be short enough, the
first term (the unit operator not changing the state) represents a dominant
contribution to the evolution, whereas the remaining term in
Eq.~(\ref{eq:evolution-free-manipulation}) takes care of the changes in the
state-vector due to the atom-atom interaction.  Because of the short storage
time, the latter term mostly couples a pair of atoms. Expanding the exponent
$e^{-i\mathcal{H}_{\mathrm{at-at}}T}$ into Taylor series we have 
\begin{equation}
e^{-i\mathcal{H}_{\mathrm{at-at}}T}-1=-iT\mathcal{H}_{\mathrm{at-at}}+\frac{1}{2}(-iT)^{2}
\mathcal{H}_{\mathrm{at-at}}^{2}+\cdots\,.\label{eq:exp-expansion}
\end{equation}
Calling on Eq.~(3) of the main text for the interaction Hamiltonian
$\mathcal{H}_{\mathrm{at-at}}$, the first term in the Taylor expansion
(\ref{eq:exp-expansion}) reads
\begin{equation}
-iT\mathcal{H}_{\mathrm{at-at}}=-i\sum_{j\neq j'}V(|\mathbf{r}_{j}-\mathbf{r}_{j'}|)T
\sigma_{ps}^{j}\sigma_{sp}^{j'}\,.
\end{equation}
The term appearing in the second order of the expansion
(\ref{eq:exp-expansion}) is 
\begin{equation}
(-iT)^{2}\mathcal{H}_{\mathrm{at-at}}^{2}=\sum_{j\neq j'}\sum_{j''\neq j'''}
V(|\mathbf{r}_{j}-\mathbf{r}_{j'}|)V(|\mathbf{r}_{j'''}-\mathbf{r}_{j''}|)\sigma_{ps}^{j}
\sigma_{sp}^{j'}\sigma_{ps}^{j''}\sigma_{sp}^{j'''}\,.
\end{equation}
When $j=j''$, $j'=j'''$ or $j=j'''$, $j'=j''$, the summation in this expression
runs only over two indices. Since $\sigma_{ps}^{j}\sigma_{ps}^{j}=0$, the
second-order term can be separated into two parts, the first part containing
double summation, the second part containing higher sums,
\begin{equation}
(-iT)^{2}\mathcal{H}_{\mathrm{at-at}}^{2}=\sum_{j\neq j'}[-iV(|\mathbf{r}_{j}
-\mathbf{r}_{j'}|)T]^{2}\sigma_{pp}^{j}\sigma_{ss}^{j'}+\mbox{nonpair terms}\,.
\label{eq:term-2-order}
\end{equation}
In a similar manner, the third-order term can be represented as
\begin{equation}
(-iT)^{3}\mathcal{H}_{\mathrm{at-at}}^{3}=\sum_{j\neq j'}[-iV(|\mathbf{r}_{j}
-\mathbf{r}_{j'}|)T]^{3}\sigma_{ps}^{j}\sigma_{sp}^{j'}+\mbox{nonpair terms}\,.
\label{eq:term-3-order}
\end{equation}
In this way, the pair summation in the cubic term contains the same operators
$\sigma_{ps}^{j}\sigma_{sp}^{j'}$ as the first-order term.  Repeating the same
procedure one arrives at the following general result for the odd and even
terms in the expansion (\ref{eq:exp-expansion}):
\begin{eqnarray}
(-iT)^{2m+1}\mathcal{H}_{\mathrm{at-at}}^{2m+1} & = & \sum_{j\neq j'}
[-iV(|\mathbf{r}_{j}-\mathbf{r}_{j'}|)T]^{2m+1}\sigma_{ps}^{j}\sigma_{sp}^{j'}
+\mbox{nonpair terms}\,,\label{eq:term-3-order-1}\\
(-iT)^{2m}\mathcal{H}_{\mathrm{at-at}}^{2m} & = & \sum_{j\neq j'}
[-iV(|\mathbf{r}_{j}-\mathbf{r}_{j'}|)T]^{2n}
\sigma_{pp}^{j}\sigma_{ss}^{j'}+\mbox{nonpair terms}\,,\label{eq:term-2n-order}
\end{eqnarray}
with $m=0,1,\cdots$. Thus, collecting in each power of the Hamiltonian
$\mathcal{H}_{\mathrm{at-at}}$ only the terms containing double sums,
Eq.~(\ref{eq:evolution-free-manipulation}) becomes
\begin{multline}
e^{-i\mathcal{H}_{\mathrm{at-at}}T}=1+\sum_{m=0}^{\infty}\frac{1}{(2m+1)!}
\sum_{j\neq j'}[-iV(|\mathbf{r}_{j}-\mathbf{r}_{j'}|)T]^{2m+1}
\sigma_{ps}^{j}\sigma_{sp}^{j'}\\
+\sum_{m=1}^{\infty}\frac{1}{(2m)!}\sum_{j\neq j'}
[-iV(|\mathbf{r}_{j}-\mathbf{r}_{j'}|)T]^{2m}\sigma_{pp}^{j}\sigma_{ss}^{j'}
+\mbox{nonpair terms}\,,
\end{multline}
where the terms that are not written explicitly contain triple and higher sums.
After summation we obtain
\begin{equation}
e^{-i\mathcal{H}_{\mathrm{at-at}}T}-1=\sum_{j\neq j'}\{\cos[V(|\mathbf{r}_{j}
-\mathbf{r}_{j'}|)T]-1\}\sigma_{pp}^{j}\sigma_{ss}^{j'}
-i\sum_{j\neq j'}\sin[V(|\mathbf{r}_{j}-\mathbf{r}_{j'}|)T]
\sigma_{ps}^{j}\sigma_{sp}^{j'}+\mbox{nonpair terms}\,.
\label{eq:expansion-two-atoms}
\end{equation}

\section{Validity of the approximation}
\label{sec:appb}

Let us consider the necessary conditions when the approximate expression
(\ref{eq:state-final-2}) for the atomic state is valid. The conditions can be
obtained by requiring the state $|\Psi(T)\rangle$ to be normalized, that is
$\langle\Psi(T)|\Psi(T)\rangle\approx1$.

Equation~(\ref{eq:state-final-2}) can be separated into two parts,
$|\Psi(T)\rangle=|\Psi_{+}\rangle+|\Delta\Psi\rangle$.  Since the initial state
is normalized, $\langle\Psi_{+}|\Psi_{+}\rangle=1$, the normalization condition
for the final state $|\Psi(T)\rangle$ reads
\begin{equation}
2\re\langle\Psi_{+}|\Delta\Psi\rangle+\langle\Delta\Psi|\Delta\Psi\rangle=0\,.
\label{eq:norm-2}
\end{equation}
Using Eq.~(\ref{eq:state-final-2}) we obtain
\begin{multline}
2\re\langle\Psi_{+}|\Delta\Psi\rangle+\langle\Delta\Psi|\Delta\Psi\rangle\approx A^{6}
\frac{\Omega_{\mathrm{p}0}^{6}}{4\Omega_{\mathrm{c}}^{6}}
\sum_{j\neq j'\neq j''}[e^{iV(|\mathbf{r}_{j}-\mathbf{r}_{j'}|)T}-1]
[e^{-iV(|\mathbf{r}_{j}-\mathbf{r}_{j''}|)T}-1]\\
+A^{8}\frac{\Omega_{\mathrm{p}0}^{8}}{16\Omega_{\mathrm{c}}^{8}}
\sum_{j\neq j'\neq j''\neq j'''}[e^{iV(|\mathbf{r}_{j}-\mathbf{r}_{j_{'}}|)T}-1]
[e^{-iV(|\mathbf{r}_{j''}-\mathbf{r}_{j'''}|)T}-1]\,.
\label{eq:tmp-1}
\end{multline}
We can estimate the expressions in Eq.~(\ref{eq:tmp-1}) replacing summation by
integration. Then we get
\begin{multline}
2\re\langle\Psi_{+}|\Delta\Psi\rangle+\langle\Delta\Psi|\Delta\Psi\rangle\approx A^{6}
\frac{\Omega_{\mathrm{p}0}^{6}}{4\Omega_{\mathrm{c}}^{6}}n^{3}
\int d\mathbf{r}\int d\mathbf{r}'[e^{iV(|\mathbf{r}-\mathbf{r}'|)T}-1]
\int d\mathbf{r}''[e^{-iV(|\mathbf{r}-\mathbf{r}''|)T}-1]\\
+A^{8}\frac{\Omega_{\mathrm{p}0}^{8}}{16\Omega_{\mathrm{c}}^{8}}n^{4}
\int d\mathbf{r}\int d\mathbf{r}'[e^{iV(|\mathbf{r}-\mathbf{r}'|)T}-1]\int d\mathbf{r}''
\int d\mathbf{r}'''[e^{-iV(|\mathbf{r}''-\mathbf{r}'''|)T}-1]\,,
\label{eq:tmp-2}
\end{multline}
where $n$ is the density of atoms. The integrals in Eq.~(\ref{eq:tmp-2}) can be
estimated as follows. Using the interaction potential $V(r)=C_{3}/r^{3}$ we
have
\begin{equation}
\int d\mathbf{r}'\left(1-\cos[V(|\mathbf{r}-\mathbf{r}'|)T]\right)=
4\pi\int_{0}^{\infty}r^{2}\left(1-\cos[V(r)T]\right)dr
=\frac{2}{3}\pi^{2}C_{3}T\equiv\frac{2}{3}\pi^{2}r_{\mathrm{c}}^{3}\,,
\label{eq:integral-1}
\end{equation}
where $r_{\mathrm{c}}=(C_{3}T)^{1/3}$ is a characteristic distance at which the
RDDI potential $V(r_{\mathrm{c}})$ becomes of the order of the inverse storage
time $T^{-1}$. On the other hand, the integral
\[
\int d\mathbf{r}'\sin[V(|\mathbf{r}-\mathbf{r}'|)T]=4\pi\int_{0}^{\infty}r^{2}\sin[V(r)T]dr
\]
does not converge at large values of $r$. To get a finite value we should take
into account a finite size of the atomic cloud. Then this integral becomes
proportional to $r_{\mathrm{c}}^{3}$. Thus the two terms in
Eq.~(\ref{eq:tmp-2}) are of the order of
$(n_{\mathrm{Ry}}r_{\mathrm{c}}^{3})^{2}(n_{\mathrm{Ry}}\mathcal{V})$ and
$(n_{\mathrm{Ry}}r_{\mathrm{c}}^{3})^{2}(n_{\mathrm{Ry}}\mathcal{V})^{2}$,
where $\mathcal{V}$ is the volume of the atomic cloud.

We can conclude, that Eq.~(\ref{eq:tmp-2}) is close to zero and the
approximation is valid when
\begin{equation}
n_{\mathrm{Ry}}r_{\mathrm{c}}^{3}\ll1
\end{equation}
and the total number $n_{\mathrm{Ry}}\mathcal{V}$ of Rydberg atoms in the
atomic cloud is not large.

\end{document}